\documentstyle[12pt]{article}
\input epsf
\begin{document}

\def\question#1{{{\marginpar{\small \sc #1}}}}
\newcommand{\eq}{\begin{equation}}
\newcommand{\en}{\end{equation}}
\newcommand{\bino}{\tilde{b}}
\newcommand{\tsquark}{\tilde{t}}
\newcommand{\gluino}{\tilde{g}}
\newcommand{\photino}{\tilde{\gamma}}
\newcommand{\wino}{\tilde{w}}
\newcommand{\gsi}{\,\raisebox{-0.13cm}{$\stackrel{\textstyle>}
{\textstyle\sim}$}\,}
\newcommand{\lsi}{\,\raisebox{-0.13cm}{$\stackrel{\textstyle<}
{\textstyle\sim}$}\,}

\rightline{RU-95-25}
\rightline{hep-ph/9508291}
\rightline{July 11, 1995}
\baselineskip=18pt
\vskip 0.4in
\begin{center}
{\bf \LARGE Phenomenology of Light Gauginos}\\
\vspace{.05in}
{\bf \LARGE I. Motivation, Masses, Lifetimes and Limits}\\
\vspace*{0.4in}
{\large Glennys R. Farrar}\footnote{Research supported in part by
NSF-PHY-94-23002} \\
\vspace{.1in}
{\it Department of Physics and Astronomy \\ Rutgers University,
Piscataway, NJ 08855, USA}\\
\end{center}
\vspace*{0.2in}{\vskip  0.1in
%\eject
%\vspace*{1.in}

{\bf Abstract:} I explore an economical variant on supersymmetric
standard models which may be indicated on cosmological grounds and
is shown to have no SUSY-CP problem.  Demanding radiative electroweak
symmetry breaking suggests that the Higgs is light; other scalar
masses may be $\sim 100-200$ GeV or less.  In this case the gluino and
photino, while massless at tree level, have 1-loop masses $m_{\gluino}
\sim 100 - 600$ MeV and $m_{\photino} \sim 100 - 1000$ MeV.  New
hadrons with mass $\sim 1 - 3 $ GeV are predicted and their lifetimes
estimated.  Existing experimental limits are discussed.
\thispagestyle{empty}
\newpage
\addtocounter{page}{-1}
%\tableofcontents
%\listoffigures
\newpage

The customary approach to studying the phenomenological implications
of supersymmetry has been to assume that the ``low energy'' effective
Lagrangian contains all possible renormalizable operators, including
in principle all possible soft supersymmetry breaking terms,
consistent with the gauge symmetries and certain global and
discrete symmetries.  Some models of SUSY-breaking naturally lead to
relations among the SUSY-breaking parameters at the scale $M_{SUSY}$,
so that, e.g., the minimal supersymmetric standard model (MSSM) requires
specification of 6-8 parameters beyond the gauge and Yukawa couplings
already determined in the MSM: $tan \beta \equiv \frac{v_U}{v_D}$, the
ratio of the two Higgs vevs; $\mu$, the coefficient of the
SUSY-invariant coupling between higgsinos; $M_0$, a universal
SUSY-breaking scalar mass; $m_{12}^2$, the SUSY-breaking mixing in the
mass-squared matrix of the Higgs scalars ($\mu B$ or $\mu M_0 B$ in
alternate notations); $M_{1,2,3}$, the SUSY-breaking gaugino masses
(proportional to one another if the MSSM is embedded in a GUT); and
$A$, the coefficient of SUSY-breaking scalar trilinear terms obtained
by replacing the fermions in the MSM Yukawa terms by their
superpartners.  To obtain predictions for the actual superparticle
spectrum in terms of these basic parameters, the renormalization group
equations for masses, mixings and couplings are evolved from the scale
$M_{SUSY}$ to the scale $m_{Z^0}$ where on account of different RG
running and flavor dependent couplings, the various scalars and
fermions can have quite different masses.  A particularly attractive
aspect of this approach is that for a heavy top quark, the
mass-squared of the Higgs field which gives mass to the charge 2/3
quarks becomes negative at low energy and the electroweak symmetry is
spontaneously broken\cite{a-gpw,ir}, with $m_{Z^0}$ a function of $A,~
M_0$ and other parameters of the theory. In this conventional
treatment of the MSSM, the lightest squark mass is constrained by
experiment to be greater than 126 GeV and the gluino mass to be
greater than 141 GeV, given favorable assumptions regarding their
decays\cite{cdf:gluinolim2}.

I will argue here that a more restrictive form of low energy SUSY
breaking may actually be used by Nature, one without dimension-3
operators, i.e., with no tree level gaugino masses or scalar trilinear
couplings.  We shall see that the remaining parameters of the theory
are rather well-constrained when electroweak symmetry breaking is
demanded, and that the resultant model is both very predictive and
consistent with laboratory and cosmological observations.  If this is
the correct structure of the low energy world, there will be many
consequences which can be discovered and investigated before the
construction of the LHC.  Some of these are discussed in a companion
paper\cite{f:102}\footnote{A preliminary discussion of many points
developed here and in (II) was given in \cite{f:99}.}, hereafter
refered to as II.  The purpose of this Letter is threefold:
\begin{enumerate}
\item Articulate the theoretical motivation for the absence of D=3
SUSY breaking operators in the MSSM.
\item Focus on the most probable portion of ($M_0$, $\mu$, $tan
\beta$) parameter space to obtain predictions for the gluino and
photino masses.
\item Determine the mass and lifetime of the lightest $R$-hadrons, and
with that information establish the experimental limits relevant to
this scenario.
\end{enumerate}

There are several reasons to suspect that there may be no dimension-3 SUSY
breaking operators in the low energy theory. Firstly, their absence
accounts for the absence of an observable neutron electric dipole
moment and other CP violating effects which arise naturally with
conventional SUSY breaking (the ``SUSY CP problem'').  With no
dimension-3 SUSY breaking operators, the only CP violating phases not
already present in the MSM are in the terms\footnote{Hatted fields
denote superfields, unhatted fields their scalar components.} $\mu
\int \hat{H_U}\hat{H_D} d^2 {\theta}$ and $m_{12}^2 H_U H_D$, and
possibly in the SUSY-breaking scalar mass-squared matrices if they are
flavor non-diagonal.  Using an $R$-transformation and a $U(1)_{PQ}$
phase rotation on the superfield $(\hat{H_U}\hat{H_D})$, phases in
both $\mu$ and $m_{12}^2$ can be removed.  Any phase which is
introduced thereby into the Yukawa terms in the superpotential can be
removed by chiral transformations on the quark superfields, merely
changing the phases which contribute to the strong CP problem (which
must be solved by some other mechanism). Since the gauge-kinetic terms
are not affected by U(1) and $R$ transformations, the preceding
manipulations do not introduce phases in interactions involving
gauginos.  Hence we see that without dimension-3 SUSY breaking, the
only phases beyond those of the MSM are in the squark mass-squared
matrices.  However to generate an electric dipole moment, which is a
chirality flip operator, requires a phase in an off-diagonal term
mixing superpartners of left and right chiral quarks.  In the case at
hand, mixing between left and right chiral superpartners is induced
only by the $\mu$ term and not also by $A$ terms as in the usual case.
Therefore the relevant mixing is real in the basis found above.  Note
that the argument given here applies to all orders of perturbation
theory, so it shows that even though the physical gluino and
neutralinos do have a mass coming from radiative corrections, no edm
is generated unless $A$ terms are present.\footnote{A discussion of
natural criteria for eliminating the SUSY CP problem in the MSSM,
including the case that scalar trilinears and gaugino masses are
present, can be found in \cite{garisto:MSSMCP}.}  Finding a neutron
edm would therefore exclude the scenario of no dimension-3 SUSY
breaking.

A second reason to consider the absence of dimension-3 SUSY breaking
terms is that they simply do not arise in many types of SUSY breaking.
The reason a distinction naturally emerges between dimension-3 and
dimension-2 SUSY breaking can be seen as follows.  A SUSY-breaking
mass for the spin-0 component of a chiral superfield $\hat{Q}$
originates from its kinetic term:
\eq
\int K(\hat{\Psi_i}^{\dagger}, \hat{\Psi_i})
\hat{Q}^{\dagger} \hat{Q} d^2 {\theta} d^2 \bar{{\theta} },
\en
where the $\Psi_i$ includes all the chiral superfields of
the theory.  The Kahler potential $K(\hat{\Psi_i}^{\dagger}
\hat{\Psi_i})$ is a vector superfield, so generally has an expansion
$1 + \frac{b_i}{M} (\hat{\Psi_i} + \hat{\Psi_i}^{\dagger}) +
\frac{c_i}{M^2} \hat{\Psi_i}^{\dagger} \hat{\Psi_i} + ... ~$.  In a
hidden sector SUSY-breaking scenario the interaction between hidden
sector and visible sector fields is purely gravitational so that $M
\sim M_{Pl}$ for terms coupling visible and hidden sector fields.
When one or more of the hidden sector superfields develops a
non-vanishing auxilliary component, a mass-squared $M_0^2 \sim
c_i (\frac{F_{\Psi_i}}{M_{Pl}})^2$ for the scalar component of
$\hat{Q}$ is produced.  On the other hand,
gaugino masses come from the superpotential:
\eq
\int f(\hat{\Psi_i}) W_{\alpha} W^{\alpha} d^2 {\theta},
\label{Fterm}
\en
where the gauge kinetic function $f(\hat{\Psi}_i)$ is a gauge singlet chiral
superfield whose expansion in hidden sector fields has the form $1 +
\frac{b_{i}}{M_{Pl}} \hat{\Psi_i}+ \frac{c_{ij}}{M_{Pl}^2}
\hat{\Psi_i} \hat{\Psi_j} + ... ~$.  If the linear terms in this
expansion have no $F-$component or are absent entirely, for example
because there are no gauge singlet hidden sector fields, then the
leading contribution to the gaugino mass is $\sim \frac{c_{ij}
F_{\Psi_i} <\Psi_j>}{M_{Pl}^2} \sim \frac{<\Psi_j>}{M_{Pl}} M_0$.  As
discussed in detail in ref.
\cite{bkn}, hidden sector models only make sense if $<\Psi'>~ <<~
M_{Pl}$, so the dimension-3 gaugino mass is negligible compared to the
dimension-2 scalar masses.  $A$ terms are produced in the same way as
the gaugino mass, replacing $ W_{\alpha} W^{\alpha}$ in eqn
(\ref{Fterm}) by the Yukawa terms, or by linear terms in the Kahler
potential after using the equation of motion to eliminate the
$F$-component of $\hat{Q}$ or $\hat{Q}^{\dagger}$.

Thus in hidden sector SUSY-breaking models in which gauge singlets do
not develop an $F$-component, the coefficients of dimension-3
operators are negligible in comparison to the coefficients of
dimension-2 operators.  More generally, this occurs whenever
linear terms which develop a non-vanishing $F$-component are absent
from the expansion of the Kahler potential, gauge kinetic function and
the analogous functions for other terms in the superpotential.  This
occurs in several models, for instance ones in which the
cosmological constant naturally vanishes in leading
order\cite{brignole_zwirner} and others in which SUSY-breaking is
driven by hidden sector gaugino condensation and the effective
Lagrangian is invariant under a phase transformation on the
condensate.

The success of standard cosmology and nucleosynthesis may be another
hint that SUSY-breaking is not driven by gauge singlet fields since
such fields generally produce severe cosmological difficulties as
shown in ref. \cite{bkn}\footnote{In special situations the
difficulties can be overcome, as shown in ref. \cite{bpr}.}.

Anticipating results to be obtained below, a final motivation for
considering this scenario is that it gives a natural explanation for
the missing matter of the Universe.  For $R$-hadron and photino masses
in the ranges predicted\cite{f:99} in this scenario, relic photinos
provide the observed level of cold dark matter in the
Universe\cite{f:100}.  In particular pions catalyze the conversion of
photinos to $R^0$'s, the gluon-gluino bound state, which annihilate
via the strong interactions.  For a critical value of $r \equiv
\frac{m(R^0)}{m_{\photino}}$ in the range $\sim 1.6 - 2.2$, the
resultant density of photinos is just what is needed.

Therefore we henceforth make the ansatz that there are no dimension-3
SUSY-breaking operators, and set all $A$'s and $M_{1,2,3}$ to zero.
The gluino and lightest neutralino, which are massless in tree
approximation, get masses at one loop from virtual top-stop pairs,
and, for the neutralinos, from ``electroweak'' loops involving
higgsino/wino/bino and Higgs/gauge bosons\cite{bgm,pierce_papa,f:96}.
The top-stop loop depends on the stop masses, especially the
splitting between the stop mass eigenstates, which is proportional to
$\mu cot \beta$ and the average stop mass.  The electroweak loops
depend on the Higgs and Higgsino masses and mixings, especially on
$\mu$, $\tan \beta$, and the masses of the heavier Higgs bosons.
These radiative corrections were estimated in ref. \cite{f:96}, in the
limit of $\mu,~M_0 >> M_Z$, assuming a common scalar mass and taking
various values of $M_0, ~\mu$, and $ tan \beta$.  There, it was
determined that in order to insure that the chargino mass is greater
than its LEP lower bound of about 45 GeV, $\mu$ must either be less
than 100 GeV (and $tan \beta {\,\raisebox{-0.13cm}{$\stackrel{\textstyle<}
{\textstyle\sim}$}\,} 2$) or greater than several TeV.

Here I will also suppose that radiative electroweak symmetry
breaking\cite{a-gpw,ir} produces the observed $Z^0$ mass for $m_{t}
\sim 175$ GeV.  This is not possible in the large $\mu$ region, so I
will consider only the low $\mu$ region: $\mu
{\,\raisebox{-0.13cm}{$\stackrel{\textstyle<}
{\textstyle\sim}$}\,} 100$ GeV.  In
addition, from Fig. 6 of ref. \cite{a-gpw} one sees that $M_0$, the
SUSY-breaking scalar mass in the Higgs sector must be $\sim 100 - 300
$ GeV, with 150 GeV being the favored value.  A more exact treatment
suggests a somewhat lower value.  Assuming that the stop mass is
similar to $M_0$, from Figs. 4 and 5 of ref. \cite{f:96} we find
$m_{\gluino} \sim 100-600$ and $m_{\photino} \sim 100-900$
MeV.\footnote{Imposing strict equality of all scalar SUSY-breaking
masses at the scale $M_{SUSY}$ is difficult or maybe impossible, since
in that case the lightest Higgs mass comes out too low given the
restrictions on $\mu, ~tan \beta$ coming from chargino and neutralino
masses\cite{lopez,diaz}.  However radiative corrections are sufficient
to give an acceptable mass to the lightest Higgs if the stop mass is
allowed to be larger than this.  This favors the low end of the gluino
mass range while not much affecting the photino mass prediction.}
Since the electroweak loop was treated in ref. \cite{f:96} with an
approximation which is valid when $M_0$ and $\mu$ are $>> m_{Z^0}$,
these results for the photino mass are only indicative of the range to
be expected.  Furthermore, in order to properly take into account the
differences between masses of various squarks and the parameter $M_0$,
a more detailed treatment is required.  For the present, I will attach
a $\sim$factor-of-two uncertainty to the electroweak loop and consider
the enlarged photino mass range $100 - 1000$ MeV.

Having outlined above the motivation for considering theories without
dimension-3 SUSY breaking operators and having focused on a
substantially restricted range of parameters, let us turn to
consideration of the most essential phenomenological properties of the
light particles of this theory.  The primary issues to be here
discussed are: {\it i)} Predicted mass and lifetime of the lightest
$R$-meson, the $g \gluino$ (glueballino) bound state denoted $R^0$.
{\it ii)} Predicted mass of the flavor singlet pseudoscalar which gets
its mass via the anomaly (the ``extra'' pseudoscalar corresponding to
the $ \gluino \gluino$ ground state degree of freedom).  {\it iii)}
The flavor singlet pseudogoldstone boson resulting from the
spontaneous breaking of the extra chiral symmetry associated with the
light gluino, which is identified as the $\eta'$.

The $R^0$ mass can be quite well determined from existing lattice QCD
calculations, as follows\cite{f:95}. If the gluino were massless and
there were no quarks in the theory (let us call this theory sYM for
super-Yang Mills), SUSY would be unbroken and the $R^0$ would be in a
degenerate supermultiplet with the lightest $0^{++}$ glueball, $G$,
and a $0^{-+}$ state I shall denote $ \tilde{\eta}$, which can be
thought of as a $\gluino \gluino$ bound state\footnote{It is
convenient to think of the states  in terms of their ``valence''
constituents but of course each carries a ``sea'' so, e.g., the
glueball may be better described as a coherent state of many soft
gluons than as a state of two gluons.  Knowledge of these aspects of
the states is not needed for estimation of their masses.}. To the
extent that quenched approximation is accurate for sYM,\footnote{The
1-loop beta function is the same for sYM as for ordinary QCD with 3
light quarks, so the accuracy estimate for quenched approximation in
ordinary QCD, $5-15\%$, should be applicable here.} the mass of the
physical $R^0$ in the continuum limit of this theory would be the same
as the mass of the $0^{++}$ glueball.  The latest quenched lattice QCD
value of $m(G)$ from the GF11 group is $1740 \pm 71$
MeV\cite{weingarten:glu1740}.  Note the increase from the $1440 \pm
110$ value given in ref. \cite{weingarten:glueballs} and used in my
earlier work\cite{f:95,f:99}.  The UKQCD collaboration
reports\cite{ukqcd:glueballs} $1550 \pm 50$ MeV for the $0^{++}$ mass,
but this error is only statistical.  Adding in quadrature a $70$ MeV
lattice error and a 15\% quenching uncertainty\footnote{The
uncertainty associated with quenched approximation with both light
quarks and gluinos was taken in \cite{f:95} to be 25\%.  However since
the estimate of the quenching error for ordinary QCD is obtained by
comparing lattice results with the hadron spectrum, it already
includes the effects of gluinos, if they are present in nature.}
leads to a total uncertainty of $\sim 270$ MeV, so I will use the
range 1.3 - 2 GeV for massless gluinos.  Experimentally, the
$f_0(1520)$ and $f_0(1720)$ seem to be the leading candidates for the
ground state glueball, but the situation is still unclear.

Physical flavor singlet $0^{++}$ states in the glueball mass region
will in general contain both glueball and $q \bar{q}$ components,
causing physical masses not to correspond to the lattice value.  While
mixing with other states causes the physical glueball and the
$\tilde{\eta}$ to shift, the $R^0$ has nothing nearby with which to
mix.  Thus the sYM glueball mass may give a better estimate of the
$R^0$ mass than it does of the physical $0^{++}$ masses.  In analogy
with the dependence of baryon mass on quark mass, we can expect
$m(R^0) \approx m(G) + m_{\gluino}$, where $m(G)$ is the unmixed
glueball mass. Therefore in view of the expected small gluino mass and
the various uncertainties discussed above, I shall adopt the estimate
$1.4 - 2.2$ GeV for the $R^0$ mass, while giving greatest credence to
the range $1.5 - 2$ GeV.\footnote{A dedicated lattice gauge theory
calculation of the masses of these particles could in principle
improve these estimates.  Such a calculation has the usual difficulty
of treating chiral fermions on the lattice, due to the Majorana nature
of the gluino.  On the other hand, since the $R^0$ does not have
vacuum quantum numbers, some of the difficulties in a glueball mass
calculation are absent.}

In sYM, which in quenched approximation is identical to ordinary QCD,
the $\tilde{\eta}$ with mass $\sim 1 \frac{1}{2}$ GeV is the
pseudoscalar that gets its mass from the anomaly.  Thus in QCD with
light gluinos the particle which gets its mass from the anomaly is too
heavy to be the $\eta'$.  However there is a non-anomalous chiral U(1)
formed from the usual chiral U(1) of the light quarks and the chiral
R-symmetry of the gluinos\cite{f:95}.  Due to the formation of $q
\bar{q}$ and $\gluino \gluino$ condensates, $<\bar{q} q>$ and
$<\bar{\lambda}\lambda>$, this chiral symmetry is spontaneously
broken.  Therefore, it is natural to identify the $\eta'$ with the
pseudogoldstone boson associated with the spontaneously broken U(1).
Using the usual PCAC and current algebra techniques, in ref.
\cite{f:95} I obtained the relationship between masses and condensates
necessary to produce the correct $\eta'$ mass (ignoring mixing):
$m_{\gluino} <\bar{\lambda}\lambda>~ \sim 10~ m_s <s \bar{s}>$.  The
required gluino condensate is reasonable, for $m_{\gluino} \sim 100 -
300$ MeV.\footnote{Note that ensuring $m_{\gluino} \gsi 100$ MeV
requires the stop squarks to be not too heavy, or else their
fractional splitting is too small given that the off-diagonal term in
the squark mass matrix, $\mu cot \beta$, is limited.  This
requires the average value of the stop mass $M_{st} $ to be
${\,\raisebox{-0.13cm}{$\stackrel{\textstyle<}
{\textstyle\sim}$}\,} 300$
GeV\cite{f:96}, which is of the same order as the value $M_0 \sim 150$
GeV indicated by electroweak symmetry breaking.  For $M_{st} = 150$
GeV, ensuring $m_{\gluino} \gsi 100$ MeV requires $\mu \gsi 40$ GeV
for $tan \beta = 2$ and $\mu \gsi 20$ GeV for $tan \beta = 1$.  This
eliminates the otherwise attractive strategy of requiring $\mu$ to
arise from SUSY-breaking.  In this scenario that would cause $\mu = 0$
due to its being a dimension-3 term.  This would solve the strong CP
problem but replace it with the old U(1) problem.}  In a
more refined discussion, the physical $\eta'$ would be treated as a
superposition of the pseudo-goldstone boson, the orthogonal state
which gets its mass from the anomaly, and the $\eta$. I have not yet
identified any clear test for the prediction that the $\eta'$ is
a pseudogoldstone boson and contains a $\sim 30 \%$ $\gluino
\gluino$ component, since model independent predictions concerning the
$\eta'$ are for ratios in which the gluino component plays no role.

An important point, independent of details of the mixing, is that
this scenario {\it predicts} the existance of a flavor singlet
pseudoscalar meson in addition to the $\eta'$ which is not a part of
the conventional QCD spectrum of quark mesons and glueballs, whose
mass should be in the $1 \frac{1}{2} - 2$ GeV range, apart from
mixing.  A detailed discussion of this and other flavor singlet
mesons will be left for the future.  Note however that the isosinglet
pseudoscalar at $1420$ MeV discovered by MarkIII\cite{mark3} and
DM2\cite{dm2} in radiative $J/\Psi$ decay and recently confirmed by
the Crystal Barrel in $p \bar{p}$ annihilation\cite{xb:1420}, is
incompatible with any conventional quark model (the closest
quark-model multiplet with an opening has a pion mass of 1800 MeV) or
glueball interpretation\cite{ukqcd:glueballs} and appears to be an
excellent candidate for the expected extra state\cite{f:93}.

Having in hand an estimate of the $R^0$ mass and photino mass, we now
return to determining the $R^0$ lifetime.  Making an absolute estimate
of the lifetime of a light hadron is always problematic.  Although the
relevant short distance operators can be accurately fixed in terms of
the parameters of the Lagrangian which we have constrained to a
considerable extent, hadronic matrix elements are difficult to
determine.  It is particularly tricky for the $R^0$ in this scenario
because the photino mass is larger than the current gluino mass and,
since $m_{\photino} \sim \frac{1}{2} m_{R^0}$, the decay is highly
suppressed even using a constituent mass for the gluino.  The decay
rate of a free gluino into a photino and massless $u \bar{u}$ and $d
\bar{d}$ pairs is known\cite{hk:taugluino}:
\eq
\Gamma_0(m_{\gluino},m_{\photino}) = \frac{\alpha \alpha_s
m^5_{\gluino}}{48 \pi M_{sq}^4} \frac{5}{9}
f(\frac{m_{\photino}}{m_{\gluino}}),
\label{GammaR0}
\en
taking $M_{sq}$ to be a common up and down squark mass.  The function
$f(y)=[(1-y^2)(1 + 2y - 7y^2 + 20y^3 - 7 y^4 + 2 y^5 + y^6) +
24y^3(1 - y + y^2)log(y)]$ contains the phase space suppression which
is important when the photino is massive.  The problem is to take
into account how interactions with the gluon and ``sea'' inside the
$R^0$ ``loans'' mass to the gluino.  If this effect is ignored one
would find the $R^0$ to be absolutely stable except for the largest
gluino and smallest photino masses.

A method of estimating the {\it maximal} effect of such a ``loan'',
and thus a lower limit on the $R^0$ lifetime, can be obtained by
elaborating a suggestion of refs. \cite{accmm,franco}.  The basic idea
is to think of the hadron (here the $R^0$) as a bare massless parton
(in this case a gluon) carrying momentum fraction $x$ and a remainder
(here, the gluino) having an effective mass $M \sqrt{1-x}$, where $M$
is the mass of the decaying hadron.  Then the structure function,
giving the probability distribution of partons of fraction $x$, also
gives the distribution of effective masses for the remainder (here,
the gluino).  Summing the decay rate for gluinos of effective mass
$m(R^0)\sqrt{1-x}$ over the probability distribution for the gluino to
have this effective mass, leads to a crude estimate or upper bound on
the rate:
\eq
\Gamma(m(R^0),z) = \Gamma_0(m(R^0),0) \int^{1-z^2}_0
(1-x)^{\frac{5}{2}} F(x) dx f(z/\sqrt{1-x}),
\en
where $z=\frac{m_{\photino}}{m(R^0)}$.  The distribution function of
the gluon in the $R^0$ is unknown, but can be bracketed with extreme
cases: the non-relativistic $F_{nr}(x) = \delta(x-\frac{1}{2})$
and the ultrarelativistic $F_{ur}(x)= 6x(1-x)$.  The normalizations
are chosen so that half the $R^0$'s momentum is carried by gluons.
Figure \ref{tau_wfn} shows the $R^0$ lifetime produced by this model,
for $M_{sq} = 150$ GeV and $m(R^0) = 1.5$ GeV, for these two structure
functions, and also for the intermediate choice $F_{10}(x) = N_{10}
x^{10}(1-x)^{10}$, as a function of $r\equiv
z^{-1}=\frac{m(R^0)}{m_{\photino}}$.  Results for any $R^0$ and
squark mass can be found from this figure using the scaling behavior
$\Gamma(m(R^0),M_{sq},z) \sim m(R^0)^{5} M_{sq}^{-4} g(z)$, as long as
it is legitimate to ignore the mass of the remnant hadronic
system, say a pion.

The decay rates produced in this model can be considered upper limits
on the actual decay rate, because the model maximizes the ``loan'' in
dynamical mass which can be made by the gluons to the gluino.  We can
get an idea of the accuracy of this model by using it to estimate the
kaon semileptonic decay rate.  $K_{\mu 3}$ decay presents a similar
dynamical problem to $R^0 \rightarrow \photino + X$ since $m_{\mu}
\sim m_s$.  (The problem is more severe for $R^0 \rightarrow \photino
+ X$ since the photino is expected to be heavier than the gluino, and
also the mass ratio $m(R^0)/m_{\photino}$ is probably less than the
ratio $m_K/m_{\mu}$.)  This model gives an approximately correct ratio
between $K_{\mu 3}$ and $K_{e 3}$ rates: 0.72 or 0.81 for the
non-relativistic and ultra-relativistic wavefunctions, respectively,
compared to the experimental value of 0.67.  However it predicts a
$K_{\mu 3}$ rates 2-4 times larger than observed, for the same two
wavefunctions, overestimating the rate as anticipated.  Since the
non-relativistic wavefunction gives better predictions for both
quantities, we will favor its predictions for the $R^0$ lifetime.

In ref. \cite{f:95} I reported the result of a comprehensive study of
relevant experiments, including all those used in the famous UA1
analysis\cite{ua1} which has widely been accepted as excluding all but
certain small ``windows'' for low gluino mass.  As noted in
\cite{f:95}, the UA1 analysis {\it assumed} that the gluino lifetime
is short enough that missing energy and beam dump experiments are
sensitive to it.  However $R$-hadrons produced in the target or beam
dump degrade in energy very rapidly due to their strong interaction
scattering length of $\sim 10-15$ cm.  Since the photino is supposed
to reinteract in the detector downstream of the beam dump or carry off
appreciable missing energy, it typically has enough energy to be
recognized only if it is emitted before the $R$-hadron interacts.  As
discussed in connection with a particular experiment in ref.
\cite{f:55}, and more generally in ref. \cite{f:95}, if the $R^0$
lifetime is longer than $\sim 5~10^{-11}$ sec this criterion is not
met and the sensitivity of beam dump and missing energy experiments
to light gluinos is degraded.

Although the $R^0$ lifetime estimate obtained above has a large
uncertainty, for nearly all of the parameter space of interest the
lower bound on the lifetime is long enough that we must deal with the
degradation issue.  By a mild theoretical idealization, we can
treat the effect of a finite lifetime analytically.  Suppose that all
$R^0$'s are produced with the same energy so that each of them has the
same time dilation factor $\gamma = \frac{E}{m(R^0)}$.  Then the ratio
of the probability of the $R^0$ producing a photino before interacting,
compared to what it would be if the lifetime were zero is:
\eq
p(\tau) = (1 + \frac{\gamma \beta c \tau}{\lambda})^{-1},
\label{taudep}
\en
taking the $R^0$ lifetime to be $\tau$ and its interaction length to
be $\lambda$ (which is approximately the same as for a nucleon, so
$\lambda \sim 10 - 15 $ cm).  The reduction in the expected number of
events when the $R^0$ lifetime is non-zero corresponds to a reduction
in sensitivity to squark mass by a factor $p(\tau)^{-1/4}$.  In the
BEBC experiment\cite{bebc}, $<\gamma> \beta c \sim 1.2 \times 10^{12}$
cm/s.  This experiment modeled the loss due to rescattering in the
dump for a given $m_{\gluino}$ and $M_{sq}$, taking $\tau$ to be the
lifetime for a free gluino to decay to a massless photino and $u
\bar{u},~d \bar{d}$ or $s \bar{s}$ pair.  This is an appropriate
procedure for the portions of parameter space in which the gluino is
much heavier than the photino and its mass is much larger than the the
confinement scale (say, $m_{\gluino} \gsi 2$ GeV).  However the
photino emission time obtained in this way is much shorter than when
the gluino is actually light and inside a massive $R^0$, and the $R^0$
lifetime is suppressed on account of the photino mass. From the BEBC
figure, their squark mass limit is $\sim 330$ GeV, for a ``gluino''
mass (effectively, $m(R^0)$)\footnote{I thank A. Cooper-Sarkar
for correspondence on this point.} of $\sim 1.7$ GeV.  This
corresponds to a lifetime of $10^{-10}$ sec using their
formula\footnote{Which gives a factor 1.8 larger lifetime than
obtained using eq. (\ref{GammaR0}) with $\alpha_s = 0.117$, since they
take $\alpha_s = 0.15$ and allow decay into $s \bar{s}$ pairs which is
kinematically forbidden in the parameter range of interest here.}.
Therefore their squark mass limit for a lifetime $10^{-9}$ ($10^{-8}$,
$10^{-7}$) sec becomes 185 GeV (107, 60) GeV rather than 330 GeV.
Note that this is essentially a limit on the mass of the lightest $u$
squark because the photino couples to charge.  The $d$ squark could be
a factor $\sim \frac{1}{\sqrt{2}}$ lighter.

In (II) I show that the experiment of Bernstein et al\cite{bernstein}
is actually insensitive to an $R^0$ in the interesting range of
masses.  Combining these new facts with the analysis of
ref. \cite{f:95} (where references are given), leads to
Fig. \ref{exptlims}, showing the excluded regions for the $R^0$
mass-lifetime plane.  ARGUS gives the light grey region, assuming
$m(R^0) = 1.5$ GeV; CUSB gives the next-to-darkest block, with its
excluded region extending over all lifetimes.  Gustafson et al gives
the next-to-lightest block in the upper portion of the figure; it
extends to infinite lifetime, but makes specific assumptions about
production rate.  UA1 gives the darkest block in the lower right
corner; it extends to higher masses and shorter lifetimes not shown on
the figure, where it is continued by collider limits.  Evidently, the
most interesting regions for the tree-level-massless gluino scenario
are essentially unconstrained by previous experiments.

The phenomenology discussed above also applies to theories with a small
tree-level gluino mass.  Compatibility with the $\eta'$ mass and the
CUSB experiment requires $100 {\rm MeV}
{\,\raisebox{-0.13cm}{$\stackrel{\textstyle<}
{\textstyle\sim}$}\,} m_{\gluino} {\,\raisebox{-0.13cm}{$\stackrel{\textstyle<}
{\textstyle\sim}$}\,} 1.5$
GeV. The photino mass would have to be tuned to be close enough to the
$R^0$ mass to avoid too much relic density in photinos\cite{f:101}.
The extent of the required tuning increases as the squark mass does.
It would be very difficult to have gluinos heavy enough to avoid the
CUSB limit, $m_{\gluino} \gsi 3.5$ GeV, while keeping the lifetime
short enough to avoid conflict with missing energy experiments.  Thus
the gluino mass must be either less than 1 GeV or greater than the
conventional limits of missing energy experiments such as ref.
\cite{cdf:gluinolim2}.

To summarize, a number of indications that dimension-3 SUSY breaking
operators may not exist in the low energy effective theory were cited.
We found that although gauginos are massless at tree level, radiative
corrections give gluino masses in the $100-300$ MeV range and photino
masses somewhat larger.  The lightest R-hadron (the ``glueballino'',
$R^0$) mass is estimated to be in the 1.4-2.2 GeV range, and its
lifetime is likely to be longer than $\sim 10^{-10}$ sec.  Therefore
beam-dump experiments are more appropriately used to provide limits on
squark masses than to exclude light gluinos.  The scenario requires
the mass of the lighter chargino to be below $m_W$, so it will be
tested at LEP.  Using signatures and detection strategies for
$R$-hadrons and squarks developed in (II), positive evidence of this
scenario could be found before that.

{\bf Acknowledgements:}  I am deeply indebted to many colleagues with
whom I have discussed or investigated various aspects of this problem,
including E. Kolb, C. Kolda, M. Luty, A. Masiero and S. Thomas.

%\appendix

%\newpage

%\bibliography{susy,qcd,f,radecay}
%\bibliographystyle{unsrt}

\begin{figure}
\epsfxsize=\hsize
\epsffile{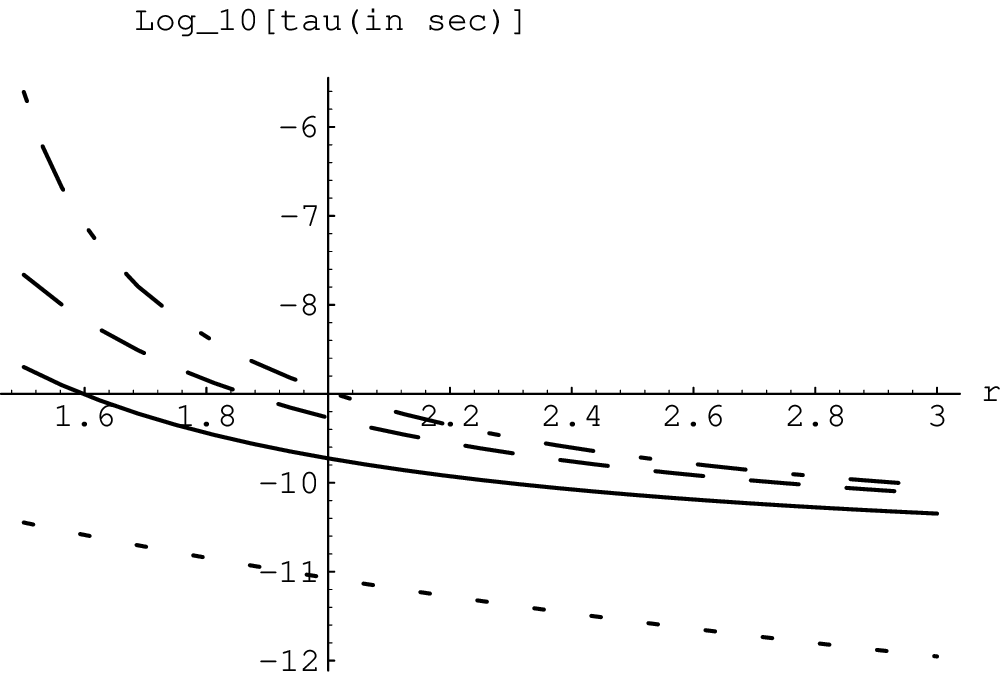}

\caption{$R^0$ lifetime in a crude model for three different gluon
distribution functions described in the text (solid: $F_{ur}$,
dashed: $F_{10}$, dot-dashed: $F_{nr}$) as a function of $r\equiv
\frac{m(R^0)}{m_{\photino}}$, with $m(R^0) = 1.5$ GeV and $M_{sq}=150$
GeV.  The dotted curve is a plot of the lifetime of a free gluino of
mass (r/1.5) GeV, decaying into massless $u \bar{u}$ or $d \bar{d}$
and$ \photino$ for $M_{sq}=150$ GeV.}
\label{tau_wfn}
\end{figure}

\begin{figure}
\epsfxsize=\hsize
\epsffile{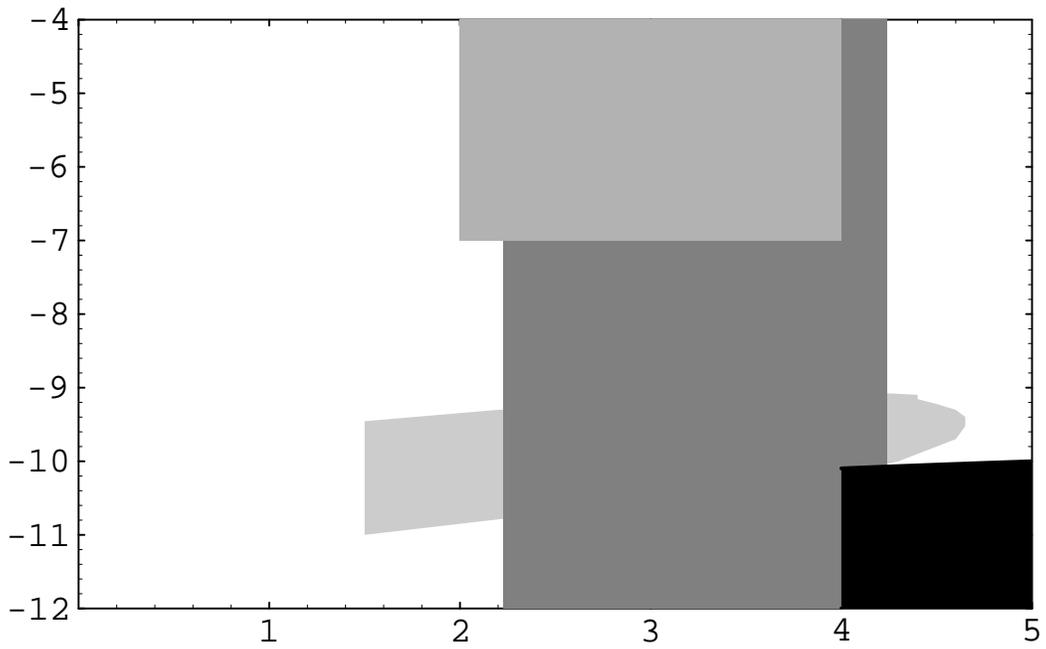}
\caption{Experimentally excluded regions of $m(R^0)$ and
$\tau_{\tilde{g}}$.  Horizontal axis is $m(R^0)$ in GeV;  vertical
axis is $Log_{10}$ of the lifetime in sec.  A massless gluino would
lead to $m(R^0) \sim 1.2-2.2$ GeV.}
\label{exptlims}
\end{figure}

\end{document}